\documentclass[a4paper,fleqn]{cas-dc}

\usepackage[authoryear]{natbib}
\usepackage{amssymb}
\usepackage{amsmath}
\usepackage{url}
\usepackage{threeparttable}
\usepackage{orcidlink}
\usepackage{float}
\usepackage[normalem]{ulem}
\usepackage{xcolor}
\restylefloat{figure} 

\newcommand{\igr}{IGR~J17014-4306}
\newcommand{\xmm}{{\it XMM}-Newton}
\newcommand{\swift}{{\it Swift}}
\newcommand{\nustar}{\textit{NuSTAR}}

\begin{document}

\let\WriteBookmarks\relax
\def\floatpagepagefraction{1}
\def\textpagefraction{.001}

\shorttitle{IGR J17014-4306: White-Dwarf Mass Constraints}
\shortauthors{D.C. Souza et~al.}

\title[mode=title]{ Constraining the White-Dwarf Mass in the Intermediate Polar \igr\ using X-Ray and Optical Spectroscopy}

\cortext[cor1]{Corresponding author. D. C. Souza E-mail: \texttt{diegocarvalhodesouza2@gmail.com}.}
\author[1]{D. C. Souza\orcidlink{0000-0001-9255-864X}}[orcid=0000-0001-9255-864X]
\cormark[1]

\author[2]{M. P. Diaz\orcidlink{0000-0002-6040-0458}}[orcid=0000-0002-6040-0458]

\author[1]{A. S. Oliveira\orcidlink{0000-0001-6422-9486}}[orcid=0000-0001-6422-9486]

\author[3,4]{G. J. M. Luna\orcidlink{0000-0002-2647-4373}}[orcid=0000-0002-2647-4373]

\author[5]{C. V. Rodrigues\orcidlink{0000-0002-9459-043X}}[orcid=0000-0002-9459-043X]

\author[6,7]{M. Reynolds\orcidlink{0000-0003-1621-9392}}[orcid=0000-0003-1621-9392]

\author[1]{N. Palivanas\orcidlink{0000-0002-6917-2842}}[orcid=0000-0002-6917-2842]

\author[8]{I. J. Lima\orcidlink{0000-0001-6013-1772}}[orcid=0000-0001-6013-1772]

\author[9]{J. Mikołajewska\orcidlink{0000-0003-3457-0020}}[orcid=0000-0003-3457-0020]

\author[10,11]{M. Orio\orcidlink{0000-0003-1563-9803}}[orcid=0000-0003-1563-9803]

\affiliation[1]{organization={Universidade Vale do Paraíba, Instituto de Pesquisa e Desenvolvimento},
               city={São José dos Campos},
               postcode={12244-000},
               state={SP},
               country={Brazil}}

\affiliation[2]{organization={Universidade de São Paulo, Instituto de Astronomia, Geofísica e Ciências Atmosféricas},
               city={São Paulo},
               postcode={05508-900},
               state={SP},
               country={Brazil}}

\affiliation[3]{organization={Universidad Nacional de Hurlingham, Laboratorio de Investigación y Desarrollo Experimental en Computación},
               city={Hurlingham},
               postcode={B1688},
               state={BA},
               country={Argentina}}

\affiliation[4]{organization={Consejo Nacional de Investigaciones Científicas y Técnicas (CONICET)},
               city={Buenos Aires},
               postcode={C1425},
               state={BA},
               country={Argentina}}

\affiliation[5]{organization={Instituto Nacional de Pesquisas Espaciais},
               city={São José dos Campos},
               postcode={12227-010},
               state={SP},
               country={Brazil}}

\affiliation[6]{organization={Department of Astronomy, University of Michigan},
               addressline={1085 South University Avenue},
               city={Ann Arbor},
               postcode={48109},
               state={MI},
               country={USA}}

\affiliation[7]{organization={Department of Astronomy, Ohio State University},
               addressline={140 West 18th Avenue},
               city={Columbus},
               postcode={43210},
               state={OH},
               country={USA}}

\affiliation[8]{organization={Universidade Estadual Paulista Júlio de Mesquita Filho (UNESP)},
               city={Guaratinguetá},
               postcode={12516-410},
               state={SP},
               country={Brazil}}

\affiliation[9]{organization={Nicolaus Copernicus Astronomical Center, Polish Academy of Sciences},
               addressline={Bartycka 18},
               city={Warsaw},
               postcode={00-716},
               country={Poland}}

\affiliation[10]{organization={University of Wisconsin-Madison, Astronomy Department},
                addressline={475 N Charter St.},
                city={Madison},
                postcode={53706},
                state={Wisconsin},
                country={USA}}

\affiliation[11]{organization={INAF-Osservatorio di Padova},
                addressline={Vicolo Osservatorio 5},
                city={Padova},
                postcode={I-35122},
                country={Italy}}

\cortext[cor1]{Corresponding author}

\begin{abstract}
Intermediate Polars (IPs) is a subclass of magnetic cataclysmic variables, often exhibit a significant discrepancy between white-dwarf (WD) mass measurements derived from X-ray spectral modeling and those from optical dynamical studies. To investigate this issue, we present a study of the IP \igr. We performed a joint spectral analysis of \nustar\ observations combined with archival \xmm\ and \swift\ data, modeling the broad-band X-ray spectrum with a multi-temperature cooling flow model. Additionally, we obtained time-resolved optical spectroscopy from the SOAR telescope to measure the radial velocity of the donor star and constrain the system's orbital parameters. The X-ray spectral analysis reveals a very high post-shock plasma temperature of $kT_{\rm max}=50_{-8}^{+11}$ keV, which corresponds to a white-dwarf mass of $M_{\text{WD}} = 0.94^{+0.09}_{-0.07}~M_{\odot}$.  We also fitted the \textit{NuSTAR} spectrum above 10~keV with the \texttt{ipolar30KK} model, obtaining $M_{\mathrm{WD}}=0.88^{+0.14}_{-0.12}\,M_{\odot}$. Our optical spectroscopy yields a radial velocity semi-amplitude for the companion star of $K_{2}=165\pm4$ km\,s$^{-1}$
and an optical mass function of $f(M)=0.249\pm0.018\,M_\odot$, which
combined with the eclipse geometry and the WD mass estimate obtained from X-ray models, favors a high mass ratio, between $0.76 \lesssim q \lesssim 0.95$, and a donor-star mass of approximately $0.67 \lesssim M_2/M_\odot \lesssim 0.89$. The extreme case of $q = 1$ results in an upper limit for the WD mass of $1.14\,M_\odot$, which, together with the X-ray measurements, indicates a relatively massive white dwarf. The lack of an independent measurement of the WD dynamical mass leaves it unclear whether the dynamical mass is consistent with the X-ray measurements or whether the X-ray mass is underestimated, as in other IPs. Thus, our constraints make \igr\ an important system in the pursuit of a better understanding of IPs.

\end{abstract}


\maketitle

\section{Introduction}
\label{introduction}

\igr\ is classified as a magnetic cataclysmic variable (mCV) of the intermediate polar (IP) subclass \citep{2013A&A...556A.120M,2017MNRAS.470.4815B}, and therefore it is understood to host a white dwarf (WD) whose magnetic field is strong enough to truncate the inner regions of the accretion disk, yet insufficient to synchronize the spin period of the WD with the orbital period. This system gathered significant attention after \cite{2017Natur.548..558S} associated it with Nova Sco 1437, a fast-declining classical nova. Those authors also identified three short-lived brightenings in the DASCH optical light curve between 1934 and 1942, and interpreted them as dwarf-nova eruptions. However, the connection of \igr\ to Nova Sco 1437 has since been disputed \citep[see e.g.][]{2019MNRAS.490.4194H,2025A&A...694A.105G}.

XMM-Newton observations of \igr\ \citep{2017MNRAS.470.4815B} show that the system experiences complete X-ray eclipses and partial eclipses in the optical B band, in addition to a modulation at the WD spin period of 1859 s. Using optical data, the orbital period was determined to be 12.8 hours, establishing \igr\ as the eclipsing IP with the longest recorded orbital period \citep{2017MNRAS.470.4815B}. Through spectral analysis of the XMM-Newton/PN and Swift/BAT data, \citet{2017MNRAS.470.4815B} estimated a WD mass of 1.16 $M_\odot$. \citet{2018MNRAS.473.4692P} observed spin-modulated circular polarization in \igr, confirming its status as an IP. The polarization, unaffected by signal changes, suggests that only one magnetic pole of the WD is visible. A decrease in circular polarization during the eclipse likely results from the cyclotron emission region being obscured. Using TESS and AAVSO eclipse timings, \citet{2025ApJ...988...87Z} claim that the orbital period of \igr\ is increasing at a very high rate, comparable to the rates found in compact binary supersoft X-ray sources (CBSS) and recurrent novae.

The accurate determination of the WD mass in mCVs is critical for the understanding of stellar evolution and the physics of accretion. However, a persistent discrepancy is often observed: masses derived from modeling of the X-ray spectra tend to be underestimated in comparison to those obtained from dynamical methods, such as radial velocity analysis, usually obtained from observations in the optical regime. This divergence, documented in systems such as YY~Dra \citep{alvarez2024mass}, GK~Per \citep{alvarez2021intermediate}, and XY~Ari \citep{alvarez2023dynamical}, highlights the need for robust dynamical measurements to calibrate and refine accretion models. Recently,\citet{2025A&A...700A.180S} investigated the systematic uncertainties affecting the X-ray spectroscopy method and proposed that this discrepancy arises preferentially in low-luminosity IPs (luminosity below  $10^{33}$ erg s$^{-1}$). In these systems, a relatively low local mass accretion might lead to an extended, tall post-shock region whose height is not accounted for in standard models, leading to underestimated masses. Conversely, IPs accreting at higher rates, where the shock is compressed against the WD surface, are not expected to present such a bias. Motivated by this scenario, we analyzed the data obtained with NuSTAR and reanalyzed archival observations from Swift and XMM-Newton in order to constrain the WD mass
in \igr. We also obtained time-resolved optical spectroscopy with the SOAR telescope to constrain some orbital parameters and provide a limit for the WD mass.

In Section~\ref{sect:obs} we present the X-ray and optical datasets and the spectra extraction procedures. Section~\ref{sec:spec} presents the results of our X-ray spectral analysis. Section~\ref{sec:mass_estimation} details the WD mass estimation from X-ray models, and section~\ref{sec:dynamical} describe how the dynamical constraints. A summary and conclusions are provided in Section~\ref{sect:disc}.

\section{Observational datasets and reduction}
\label{sect:obs}

\subsection{X-ray observations}

\subsubsection{\nustar}

\textit{NuSTAR} observed \igr\ on July 1st 2017, with an exposure time of 37.1~ks in both focal plane modules, FPMA and FPMB (ObsID 30301024002). Data reduction was carried out with the \texttt{nupipeline} tool, which applies calibration and screening to generate cleaned event files. The spectral products were generated using the \texttt{nuproducts} task. For spectral extraction, we defined a circular source region with a radius of 70~arcseconds centered on the SIMBAD source coordinates ($\alpha$=17\textsuperscript{h}01\textsuperscript{m}28.15\textsuperscript{s}, $\delta$=$-43^{\circ}06^{\prime}12^{\prime\prime}$), and a background region of equal size located on the coordinates $\alpha$=$17^{\rm h}01^{\rm m}10^{\rm s}$, $\delta$=$-43^{\circ}08^{\prime}26^{\prime\prime}$, in an adjacent area free of contaminating sources on the same detector chip as \igr. The resulting spectra were grouped to ensure a minimum of 25 counts per bin, allowing the use of $\chi^2$ statistics in subsequent spectral modeling.

\subsubsection{XMM-Newton}

\igr\ was observed with \textit{XMM-Newton} (ObsID 0761120701), using the EPIC camera. These data were already presented by \citet{2017MNRAS.470.4815B}, but here we extracted the spectra using the latest calibration set and following standard procedures. We remark that our newly extracted products do not differ significantly from those presented by \citet{2017MNRAS.470.4815B}.

\subsubsection{Swift}

We analyzed observations of \igr\ obtained with the X-Ray Telescope (XRT; \citealt{Burrows2005}) onboard the \textit{Neil Gehrels Swift Observatory} using the Photon Counting (PC) readout mode. The spectrum was generated using the online \textit{Swift-XRT Product Generator}\footnote{\url{https://www.swift.ac.uk/user_objects/}} provided by the UK Swift Science Data Center \citep{Evans2007}.

\subsection{Optical spectroscopic data}

\begin{figure}[htb]
  \centering
  \includegraphics[width=\columnwidth]{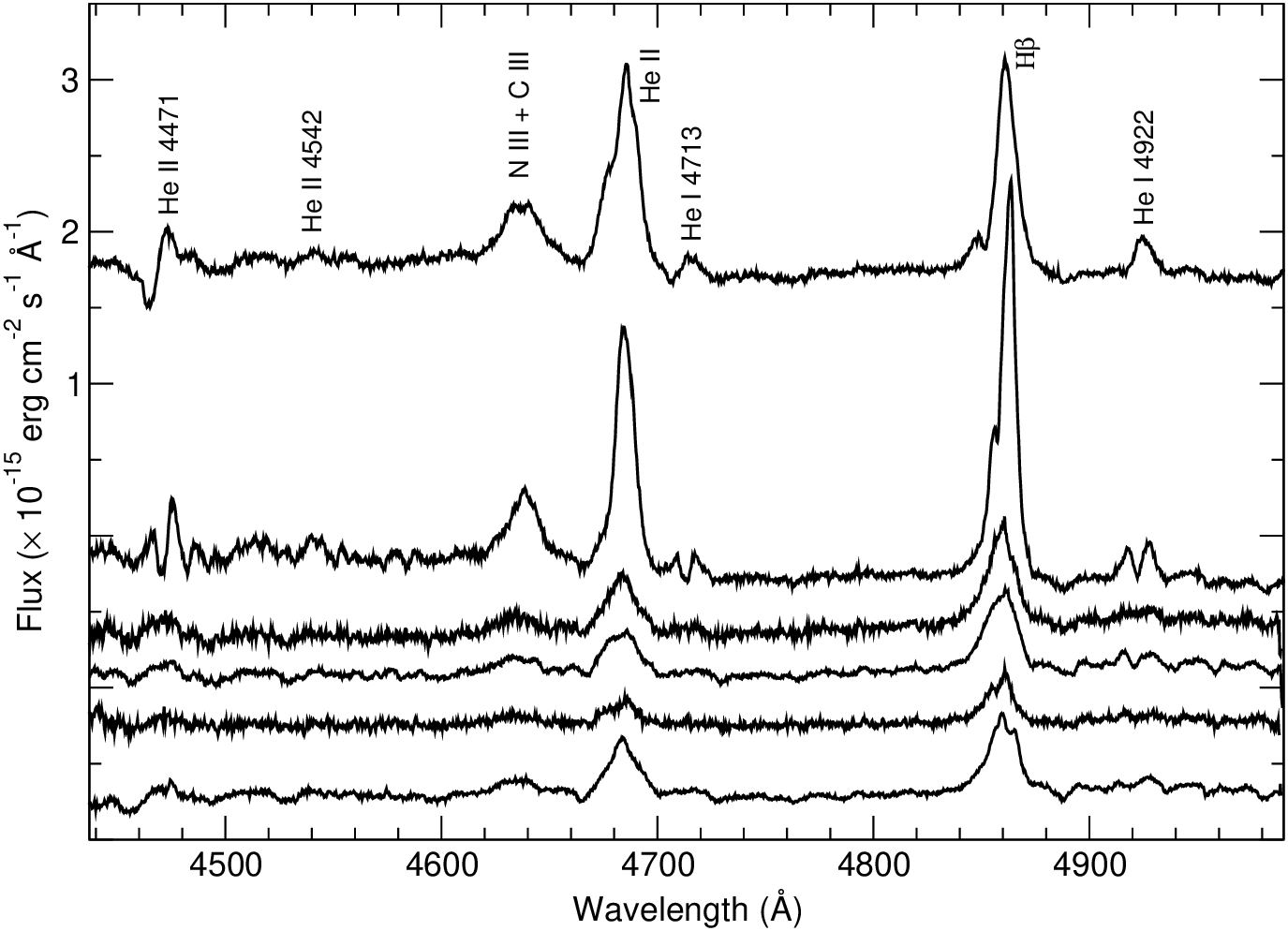}
  \caption{Nightly averaged SOAR optical spectra of \igr. The flux scale applies to the upper spectrum, which is the average from 2022 May 26. The subsequent spectra (from top to bottom) are averages from 2022 May 27, 2023 May 19, 20, 21 and 24, which were vertically shifted arbitrarily for clarity.}
  \label{fig:SOARspecs}
\end{figure}

Optical spectroscopic observations of \igr\ were conducted using the Goodman spectrograph mounted on the SOAR telescope for six nights in May 2022 and May 2023 (Figure~\ref{fig:SOARspecs}). The data set comprises 121 exposures, each with an integration time of 1200 s. The observations were carried out using the blue camera, the 2400 l/mm grating, and a 0.8$^{\prime\prime}$ slit, covering the spectral range from 4350 to 5100~\AA\ and providing a spectral resolution of 0.7~\AA\ FWHM. Bias and flat-field frames were acquired for CCD calibration, and CuHeAr arc lamps were taken at intervals shorter than one hour to ensure accurate wavelength calibration throughout the time series. The Atmospheric Dispersion Corrector (ADC) was used to minimize flux losses due to differential atmospheric refraction. A standard spectrophotometric star was observed each night with the same instrumental setup for flux calibration. The SOAR data were reduced using the IRAF\footnote{NOIRLab IRAF is distributed by the Community Science and Data Center at NSF NOIRLab, which is managed by the Association of Universities for Research in Astronomy (AURA) under a cooperative agreement with the U.S. National Science Foundation.} package following standard procedures.

\section{X-ray spectral analysis}
\label{sec:spec}

We performed a joint spectral analysis of the X-ray data from \textit{XMM-Newton} $pn$, $mos1$ and $mos2$ cameras, \textit{Swift}, and \textit{NuSTAR} using version 12.12 of the \texttt{XSPEC} package \citep{arnaud1996}. 

\begin{figure}[thb]
  \centering
  \includegraphics[trim={1.8cm 2.3cm 2.3cm 1.8cm}, clip,
                   width=\columnwidth, height=\columnwidth]{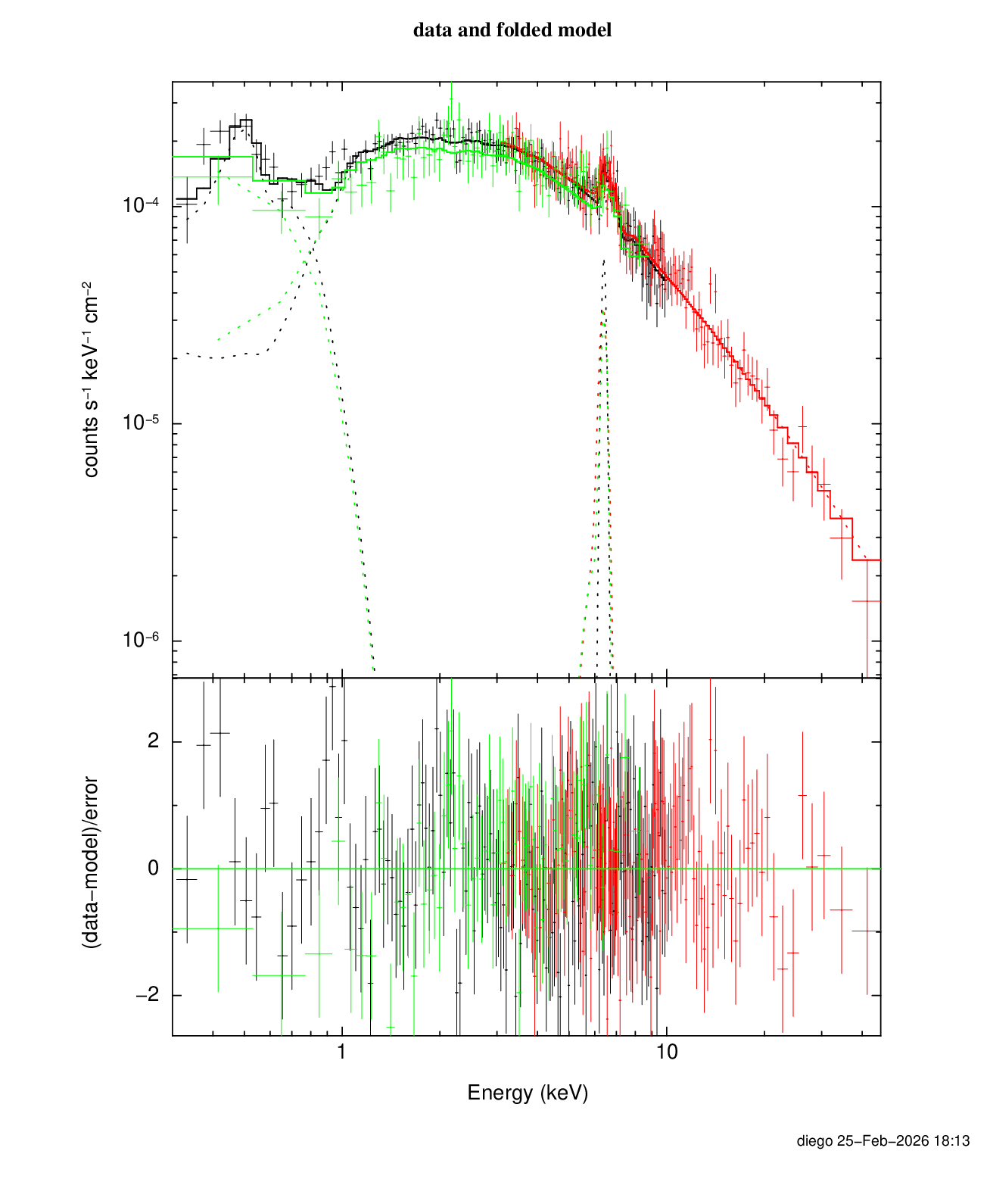}
  \caption{\textit{XMM-Newton} (we show only the $pn$-camera (black) for clarity), \textit{Swift} (green), and \textit{NuSTAR} (red) X-ray spectra of \igr\ together with the best-fit model (solid lines). The model consists of an absorbed cooling flow plasma (\texttt{mkcflow}), plus a soft blackbody (\texttt{bbody}), and a Gaussian line representing the Fe~K$\alpha$ fluorescence at 6.4 keV. The residuals of the fit are displayed in the lower panel.}
  \label{fig:modeligr}
\end{figure}

The X-ray spectra were modeled with a composite model consisting of an absorbed, multi-temperature cooling flow plasma, a soft blackbody component, and a Gaussian profile for the Fe~K$\alpha$ fluorescence at 6.4 keV. The absorption includes both a component for the interstellar medium and a partial covering absorber to account for intrinsic obscuration. The primary emission was described by a cooling flow model (\texttt{mkcflow}), representing the post-shock, optically thin plasma. The model also incorporates a soft blackbody component (\texttt{bbody}), likely originating on the WD surface. The complete model provides a good fit to the data, with $\chi^2 = 320.12$ for $340$ degrees of freedom (d.o.f.), resulting in $\chi^2_{\nu} = 0.94$. The best-fit parameters for all components are detailed in Table~\ref{tab:model_xspec}. Figure~\ref{fig:modeligr} shows the spectra from \textit{XMM-Newton}, \textit{Swift}, and \textit{NuSTAR}, along with the best-fit model.

The high value of the maximum plasma temperature ($kT_{\rm max} = 50_{-9}^{+11}$ keV) confirms the presence of a strong shock front, characteristic of accretion onto a magnetic WD. The \texttt{mkcflow} model, together with the Gaia distance ($988_{-36}^{+37}$ pc), allowed us to estimate the mass accretion rate to be $\dot{M} = 2.9 \times 10^{-10}\,M_\odot\,\mathrm{yr^{-1}}$. Our single cooling flow model provides a constrained shock temperature that is consistent with the lower limit of the hotter plasma in the two-temperature \texttt{mekal} model of \citet{2017MNRAS.470.4815B} ($kT_{\rm h} \geq 54$ keV). Importantly, both our analysis and that of \citealt{2017MNRAS.470.4815B} confirm the presence of a soft blackbody component, with our model finding a temperature of $kT_{\rm bb} = 59 \pm 3$~eV while \citet{2017MNRAS.470.4815B} found $53 \pm 1$~eV.


\renewcommand{\arraystretch}{1.3}
\begin{table*}
\centering
\begin{threeparttable}
\caption{\parbox{0.65\linewidth}{Best-fit parameters of the spectral model applied to the X-ray data, including derived physical quantities. Uncertainties are quoted at the $1\sigma$ confidence level. The normalization of the \texttt{gaussian} component is given in units of  $\mathrm{photons~cm^{-2}~s^{-1}}$.}}
\begin{tabular}{l l l}
\hline
Component & Parameter & Best-fit Value \\
\hline
\texttt{TBabs} (interstellar)      & Column density $N_{\mathrm{H}}$          & $0.79_{-0.11}^{+0.10} \times 10^{22}~\mathrm{cm}^{-2}$ \\
\texttt{partcov*TBabs} (intrinsic) & Column density $N_{\mathrm{H}}$          & $7.2_{-1.7}^{+2.0} \times 10^{22}~\mathrm{cm}^{-2}$    \\
                                   & Covering fraction                         & $0.61 \pm 0.04$                                         \\
\texttt{mkcflow}                   & Maximum temperature ($kT_{\mathrm{max}}$) & $50_{-9}^{+11}~\mathrm{keV}$                            \\
                                   & Abundance ($Z/Z_{\odot}$)                & $0.56_{-0.16}^{+0.19}$                                  \\
                                   & Mass accretion rate ($\dot{M}$)        & $2.9 \times 10^{-10}\,M_\odot\,\mathrm{yr^{-1}}$       \\
\texttt{gaussian}                  & Fe line at $6.4~\mathrm{keV}$            & $1.6 \pm 0.4 \times 10^{-5}$                            \\
\texttt{bbody}                     & Temperature $kT$                         & $59 \pm 3~\mathrm{eV}$                                  \\
                                   & Emitting region radius                  & $147 \pm 19~\mathrm{km}$                                \\
\hline
\end{tabular}
\label{tab:model_xspec}
\end{threeparttable}
\end{table*}
\section{White-dwarf mass estimation}
\label{sec:mass_estimation}

Here we present the WD mass estimate derived from modeling the X-ray spectrum.

\subsection{The post-shock temperature}\label{pst}

Following the methodology for magnetic cataclysmic variables, we can estimate the WD mass from the maximum temperature of the post-shock plasma ($kT_{\rm max}$). This temperature is directly related to the gravitational potential well of the WD, as described by \citet{Aizu1973}. To solve for the mass, we combined this relation with the theoretical mass-radius relationship for WDs \citep{Nauenberg1972}, following {  equation 3} as presented by \citet{yu2018empirical} which is based on methods established in earlier works \citep[e.g.,][]{suleimanov2005rxte,yuasa2010white}. Using our best-fit maximum shock temperature of $kT_{\rm max} = 50_{-9}^{+11}$ keV, we solved for the mass and found $M_{\rm WD} = 0.94^{+0.09}_{-0.07}\,M_{\odot}$.

\subsection{The WD mass from the \texttt{IPOLAR30KK} model}

We used the \texttt{ipolar30KK}\footnote{https://heasarc.gsfc.nasa.gov/docs/software/xspec/models/ipolar30KK.html} \citep{2016A&A...591A..35S, 2025A&A...700A.180S}  model in \texttt{XSPEC} to fit the \textit{NuSTAR} spectrum. This model adopts a physical approach to the the post-shock accretion column and includes the WD mass as a parameter. The \textit{NuSTAR} data were restricted to energies above 10~keV because the lower-energy range is affected by photoelectric absorption, which can introduce degeneracies between the absorbing column density and the intrinsic continuum.  The high-energy turnover associated with the maximum post-shock temperature is not pronounced within the fitted energy range, limiting the direct spectral constraint on the WD mass. The reference mass adopted for the calculation of the magnetospheric radius, $M_{\mathrm{WD}} = 0.94\,M_{\odot}$, was obtained from the spectral analysis presented in Section~\ref{pst}. The WD radius was calculated from the theoretical mass--radius relation of Nauenberg \citep{Nauenberg1972}, giving $R_{\mathrm{WD}} = 5.92 \times 10^{8}$~cm.

Using the WD spin period, $P_{\mathrm{spin}} = 1859$~s, we calculated the corotation radius as
\begin{equation}
{
R_{\mathrm{co}} =
\left(
\frac{G M_{\mathrm{WD}} P_{\mathrm{spin}}^{2}}
{4\pi^{2}}
\right)^{1/3},
}
\end{equation}
where $R_{\mathrm{co}}$ is the corotation radius, $G$ is the gravitational constant, $M_{\mathrm{WD}}$ is the WD mass, and $P_{\mathrm{spin}}$ is the WD spin period. For the adopted parameters this equation gives $R_{\mathrm{co}} = 2.22 \times 10^{10}$~cm. Following \citet{2025A&A...700A.180S}, we adopted $R_{\mathrm{m}} = 0.75R_{\mathrm{co}}$, corresponding to $R_{\mathrm{m}} = 1.67 \times 10^{10}$~cm and $R_{\mathrm{m}}/R_{\mathrm{WD}} = 28.2$. This value was fixed during the \texttt{ipolar30KK} fit, since the magnetospheric radius was not independently constrained by the X-ray spectrum.

\begin{figure}[tbh]
  \centering
  \includegraphics[trim={1.4cm 1.35cm 0.8cm 2.15cm}, clip, angle=-90,
                   width=\columnwidth ]{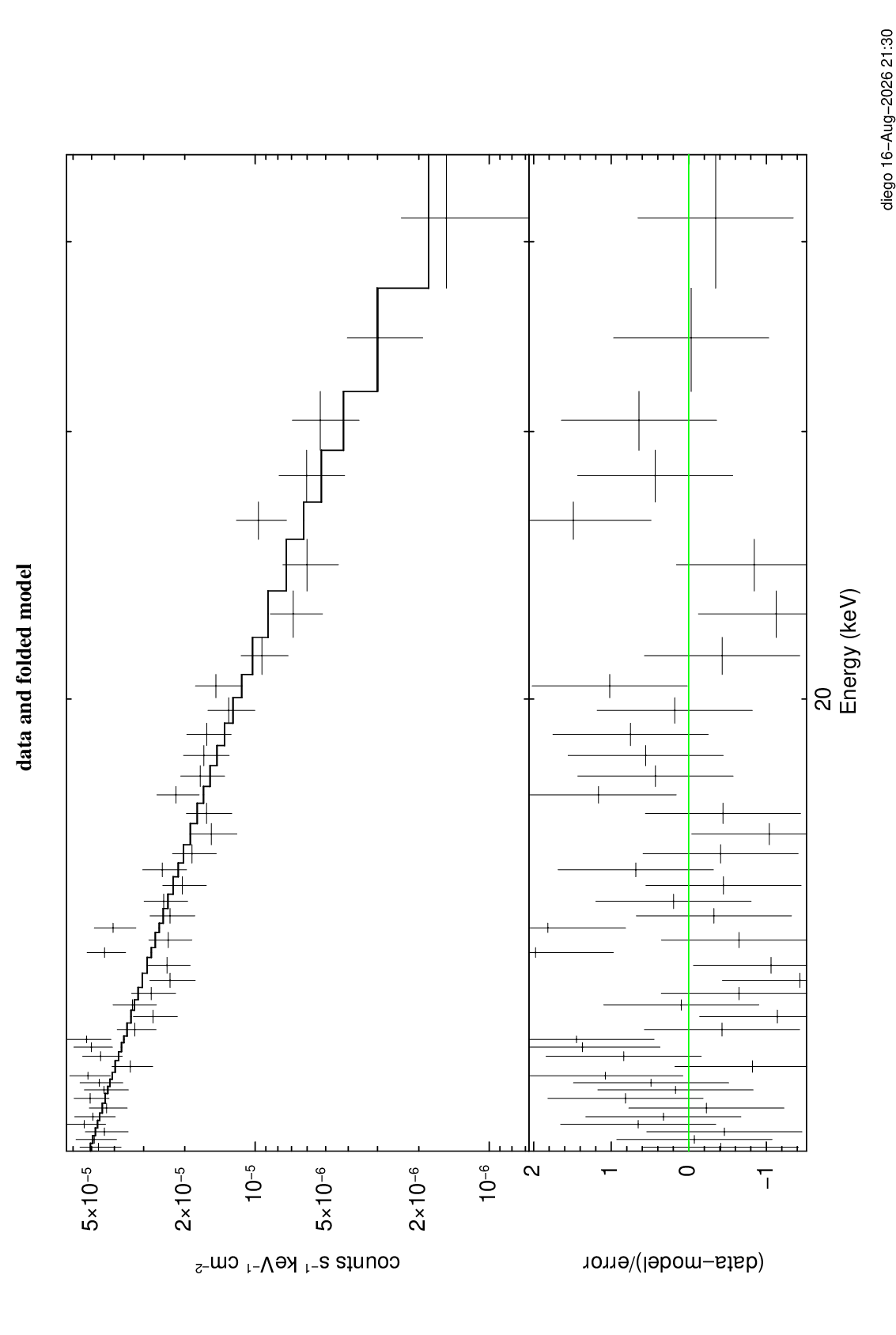}
  \caption{\textit{NuSTAR} spectrum of \igr\ fitted with the \texttt{ipolar30KK} model. Only data above 10~keV were included in the fit to avoid the region where the photoelectric absorption affects the spectrum. The data and the folded best-fit model are shown in the upper panel, while the lower panel displays the residuals. The ratio $R_{\mathrm{m}}/R_{\mathrm{WD}}=28.2$ was fixed during the fit. The best-fit WD mass is $M_{\mathrm{WD}}=0.88_{-0.12}^{+0.14}\,M_{\odot}$.}
  \label{fig:ipolar}
\end{figure}

The best fit has a $M_{\mathrm{WD}} = 0.88_{-0.12}^{+0.14}\,M_{\odot}$, with $\chi^{2} = 31.97$ for 42 degrees of freedom ($\chi^2_{\nu} = 0.76$). The corresponding spectral fit is shown in Figure \ref{fig:ipolar} and the results are presented in Table \ref{tab:model_ipolar}.

\begin{table*}[b]
\centering
\caption{Best-fit parameters and derived quantities for the \texttt{ipolar30KK} model applied to the \textit{NuSTAR} spectrum.}
\label{tab:model_ipolar}
\begin{threeparttable}
\begin{tabular}{l l l}
\hline
Component & Parameter & Best-fit Value \\
\hline

\texttt{ipolar30KK}
& WD mass ($M_{\mathrm{WD}}$)
& $0.88_{-0.12}^{+0.14}\,M_{\odot}$ \\

& Magnetospheric-radius ratio
  ($R_{\mathrm{m}}/R_{\mathrm{WD}}$)
& $28.2$ (fixed) \\

\hline
\end{tabular}
\end{threeparttable}
\end{table*}

\section{Dynamical constraints}
\label{sec:dynamical}

The H$\beta$ and He~II emission lines in our optical spectra (Figure~\ref{fig:SOARspecs}) are unsuitable for tracing the WD's motion. They are affected by self-absorption, nebular contamination, blending with absorption features, and, in the case of He~II, contamination from the nearby Bowen blend. The low signal-to-noise ratio prevents these components from being reliably separated. Moreover, radial velocities measured from the H$\beta$ wings using the method of \citet{1980ApJ...238..946S} yield non-sinusoidal and erratic curves. We therefore used the absorption-line system of the secondary star to derive the radial-velocity curve, as described below.

The optical continuum between emission lines in \igr\ shows absorption lines with depth well above the noise level of the phase-binned spectra. A late-type companion spectrum was identified by \citet{2017Natur.548..558S}, who estimated a companion of spectral type K3 from the absorption line system. Our observations are consistent with such a spectral type. Unfortunately, the limited spectral coverage and S/N do not allow discrimination between a dwarf and a giant spectrum. By selecting spectral regions between the emission lines and using an averaged template from the target itself, relative cross-correlation radial velocities were derived using 12 phase-binned spectra with the IRAF package \texttt{FXCOR}. The target template is wavelength-masked from a single phase bin spectrum, selected by its S/N ratio. The orbital ephemeris with a period of $0.5340263(5)$~d from \citet{2017Natur.548..558S} was adopted, which is precise enough for phasing all data within the observation time frame.

\begin{figure}[]
  \centering
  \includegraphics[width=\columnwidth]{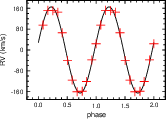}
  \caption{Radial velocity curve for the absorption lines in \igr. The sinusoidal fit (black line) yields a semi-amplitude of $K_2 = 165 \pm 4$~km\,s$^{-1}$. Individual velocity uncertainties are $\sim$15~km\,s$^{-1}$.}
  \label{fig:rvcurve}
\end{figure}

Although no systemic velocity can be derived with this method, a better match is achieved by using the absorption spectrum of \igr\ as template, yielding lower uncertainties in measured radial velocities of $\sim 15$~km\,s$^{-1}$ (Figure~\ref{fig:rvcurve}). An observed semi-amplitude $K_2 = 165 \pm 4$~km\,s$^{-1}$ for the companion absorption lines was found assuming a circular orbit. Using this value and the orbital period, the corresponding mass function is $f(M) = 0.249 \pm 0.018\,M_\odot$.

The companion irradiation and its associated impact on the absorption line depths is unknown (e.g., \citealt{WadeHorne1988}). If significant, this may lead to a smaller semi-amplitude for the secondary center of mass, as the illuminated hemisphere eventually presents veiled or no absorption lines at all. On the other hand, one may speculate that the high-energy emission due to magnetic accretion is shielded by the accretion disk and that the large binary separation also diminishes the secondary irradiation.

\begin{figure}[]
  \centering
  \includegraphics[width=\columnwidth]{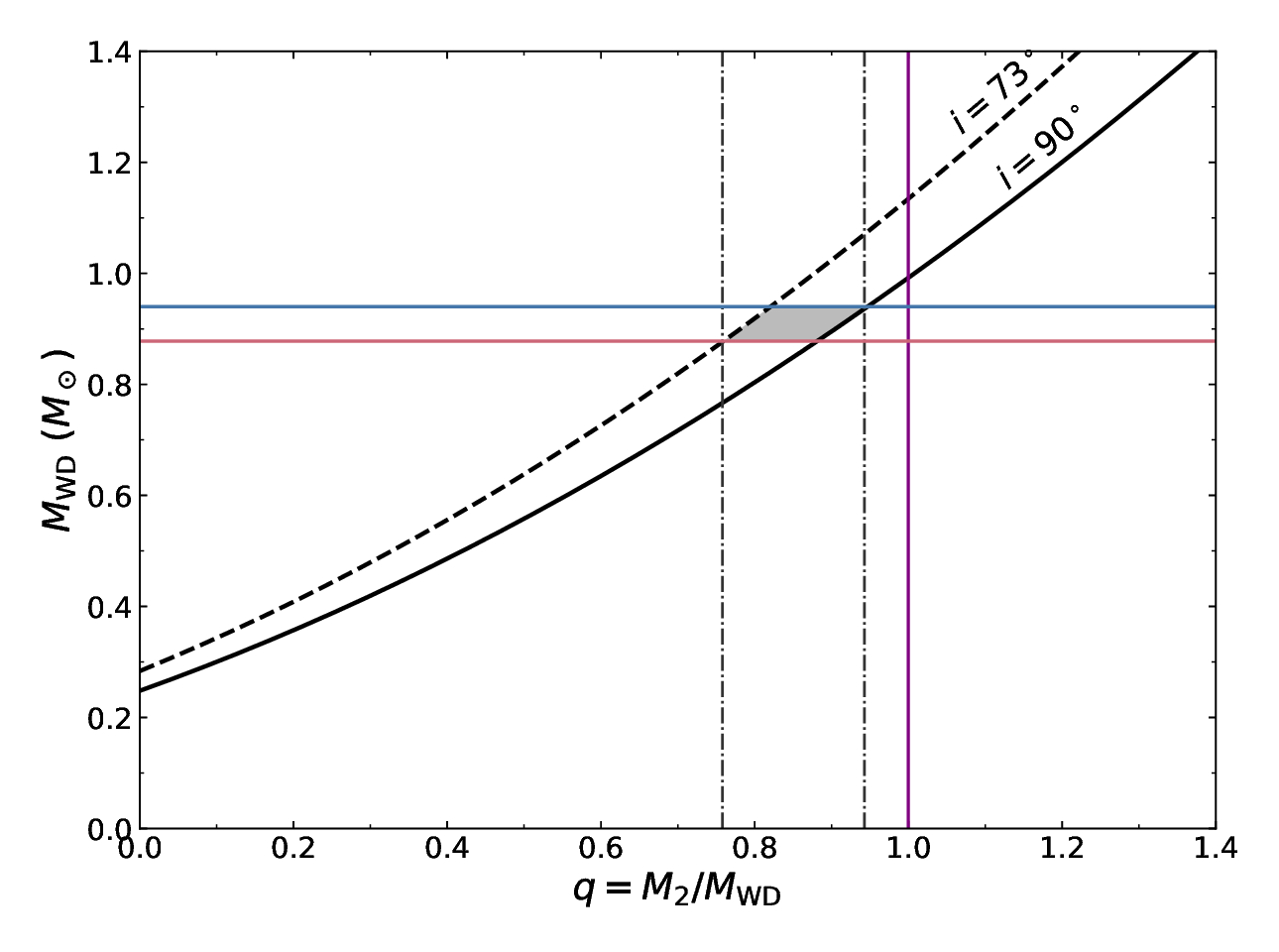}
  \caption{Constraints on the  $M_{\rm WD}$ and  $q$ of \igr. The solid and dashed black curves show the optical mass-function relations for $i=90^{\circ}$ and $i=73^{\circ}$, respectively. The horizontal blue line indicates the WD mass inferred from the shock temperature, while the red one indicates the result from the \texttt{IPOLAR30KK} model. The vertical black dashed lines mark the corresponding intersections with the mass-function relations and hence the allowed limits on $q$. The shaded region represents the range of $q$ constrained by the optical data, while the purple vertical line indicates the conditional upper limit on $M_{\rm WD}$ from the optical mass function.}
  \label{fig:mass}
\end{figure}

In the case of an eclipsing single-lined system, one may constrain the stellar masses under the assumption that the companion follows the main-sequence mass--radius relation and fills its Roche lobe. However, considering the long orbital period ($>12$~h), the companion in this system is likely evolved. A plausible range of inclinations between $73^{\circ}$ and $90^{\circ}$ (for $q = M_2/M_1 < 1.0$ and a WD eclipse width $\Delta\phi > 0.08$; \citealt{Horne1985}), combined with the optical mass function, defines a family of solutions in the $M_{\rm WD}$--$q$ plane. By adopting the white-dwarf masses determined from the X-ray models as independent constraints, we constrain the mass ratio to $0.76 \lesssim q \lesssim 0.95$. These values correspond to donor-star masses of approximately $0.67 \lesssim M_2/M_\odot \lesssim 0.89$. The condition $q<1$, can be used to define a conditional upper limit on the white-dwarf mass from the optical mass function. For $q=1$ and $i=73^{\circ}$, the optical mass function gives $M_{\rm WD}\simeq1.14\,M_\odot$, and therefore optical $M_{\rm WD}<1.14\,M_\odot$ within the adopted optical and geometrical constraints. Figure~\ref{fig:mass} shows these constraints in the $M_{\rm WD}$--$q$ plane.

Nonetheless, measuring reliable $K_1$ values from tomographic studies of the emission line wings and narrowing the orbital inclination from the WD eclipse ingress and egress features would be essential to complete the binary dynamical solution.


\section{Summary and conclusions}
\label{sect:disc}

We present a combined X-ray and optical analysis of the intermediate polar \igr, yielding a self-consistent mass determination for its WD. Our analysis of \nustar, \xmm, and \swift\ data reveals a post-shock temperature of $kT_{\mathrm{max}} = 50^{+11}_{-9}$~keV, which implies a WD mass of $M_{\rm WD} = 0.94^{+0.09}_{-0.07}\,M_{\odot}$. 
We also fitted the \textit{NuSTAR} spectrum above 10~keV with the \texttt{ipolar30KK} model. This model provides an acceptable statistical description of the spectrum and yields $M_{\mathrm{WD}} = 0.88^{+0.14}_{-0.12}\,M_{\odot}$.  The mass derived from the \texttt{ipolar30KK} fit should  be interpreted as the result of a physically motivated model that is consistent with the data, but whose precision is limited by the absence of a clearly detected high-energy cutoff.

The results for \igr\ are also consistent with the systematic effects discussed by \citet{2025A&A...700A.180S}. Using \textit{Swift}/BAT data, they obtained $M_{\mathrm{WD}} = 0.91 \pm 0.21\,M_{\odot}$, in excellent agreement with our X-ray mass estimate. Our estimate, obtained with the \texttt{ipolar30KK} model, has smaller statistical uncertainties than their reported value.

The optical spectroscopy yielded a semi-amplitude of $K_2=165\pm4$ km s$^{-1}$, resulting in a mass function of $f(M)=0.249\pm0.018\,M_\odot$, which, combined with the eclipse geometry and the white-dwarf mass estimates obtained from X-ray models, favors a high mass ratio, $0.76\lesssim q\lesssim0.95$, close to the stable mass-transfer limit $q=1$. These results imply a donor star with a mass of betwen $0.67\lesssim M_2/M_\odot\lesssim0.89$. These constraints differ from those reported by \citet{2017Natur.548..558S}, who obtained mass functions of $f(M)=0.590^{+0.245}_{-0.166}\,M_\odot$ and $0.974^{+0.105}_{-0.055}\,M_\odot$, based on larger radial-velocity amplitudes derived from a sample of only four radial-velocity data points. Assuming a white-dwarf mass between $1.0$ and $1.4\,M_\odot$, their values implied donor masses ranging from $0.3$ to $0.76\,M_\odot$ for the first solution and below $0.2\,M_\odot$ for the second. Thus, the dynamical constraints obtained here do not reproduce the earlier estimates and instead indicate a higher mass ratio and likely a more massive donor-star. The limiting case, adopting $q=1$ and $i=73^\circ$, yields a upper-limit white-dwarf mass of $M_{\rm WD}\simeq1.14\,M_\odot$, providing the conditional upper limit $M_{\rm WD}<1.14\,M_\odot$.

This result is particularly relevant in the context of IPs, where WD masses determined by X-ray shock temperatures are often underestimated compared to dynamically measured masses, as recently demonstrated for GK~Per \citep{alvarez2021intermediate}, XY~Ari \citep{alvarez2023dynamical}, and especially YY~Dra \citep{alvarez2024mass} and EX~Hya \citep[see e.g.][]{Beuermann2024}.

Although our optical data do not allow for an independent measurement of the white-dwarf mass, the optical upper limit lies above our X-ray estimates, leaving open the possibility that \igr\ shares the discrepancy reported for other IPs. This is consistent with the shock-height effect, whereby an extended post-shock region in low-luminosity IPs can lead to an underestimation of the white-dwarf mass from X-ray spectroscopy. However, due to the fact that this result is an upper limit, our analysis does not establish such an offset. A more direct test will require an independent dynamical determination of the white-dwarf mass.

In addition to the shock-height effect,
several mechanisms may bias $kT_{\rm max}$ to low values, including complex absorption and reflection, departures from the simple strong-shock free-fall model, and additional cooling via cyclotron emission. So, an apparently good spectral fit does not guarantee an accurate mass determination. Cyclotron cooling is, however, expected to be irrelevant for relatively massive WDs \citep[$M_{\rm WD} \gtrsim 0.8\,M_\odot$;][]{cropper1999effects}, where bremsstrahlung generally dominates the post-shock energy losses. In \igr, the broad X-ray bandpass, explicit modeling of partial covering absorption, and the circular polarization level ($\lesssim 4\%$; \citealt{2018MNRAS.473.4692P}) all suggest that bremsstrahlung cooling indeed dominates and that cyclotron cooling does not strongly depress the shock temperature, making the standard $kT_{\rm max}$-mass relation more reliable. 

Although our optical data do not allow a precise determination of the white-dwarf mass, they provide important constraints on the binary system that complement the X-ray analysis, making \igr\ an important system for improving our understanding of white-dwarf mass determinations in IPs.


\section*{Acknowledgements}

The authors thank an anonymous referee by their valuable comments.
DCS thanks the Coordenação de Aperfeiçoamento de Pessoal de Nível Superior (CAPES). GJML is a member of the CIC-CONICET (Argentina). MPD thanks the Conselho Nacional de Desenvolvimento Cient\'{\i}fico e Tecnol\'ogico (CNPq) under grant 310309. CVR thanks the Brazilian Space Agency (AEB) by the support from PO 20VB.0009 and the Brazilian National Council for Scientific and Technological Development -- CNPq (Proc: 305991/2024-8). JM is a member of Nicolaus Copernicus Astronomical Center, Polish Academy of Sciences and was supported by the Polish National Science Centre (NCN) grant 2023/48/Q/ST9/00138. IJL thanks \textit{São Paulo Research Foundation} (FAPESP) for financial support under grant \#2024/14358-9 and \#2024/03736-2. This research has made use of data obtained with the \nustar\ mission, a project led by the California Institute of Technology (Caltech), managed by the Jet Propulsion Laboratory (JPL) and funded by NASA. This research has made use of the \nustar\ Data Analysis Software (NuSTARDAS) jointly developed by the ASI Science Data Center (ASDC, Italy) and the California Institute of Technology (Caltech, USA). We acknowledge use of the NASA HEASARC \swift\ tools as well as Astropy.

\appendix

\printcredits

\bibliographystyle{cas-model2-names}
\bibliography{references-report2}

\end{document}